\documentclass[12pt]{article}
\usepackage{amsfonts,graphicx,amsmath,amsthm,amssymb,epsfig}
\usepackage{float,xcolor}
\allowdisplaybreaks

\begin{document}

\title{\bf Implications of Rastall Theory on Stellar Solutions admitting Vanishing Complexity:
A New Perspective}
%Imprints of Vanishing Complexity in Self-gravitating Stellar Models
\author{Tayyab Naseer$^{1,2}$ \thanks{tayyab.naseer@math.uol.edu.pk; tayyabnaseer48@yahoo.com}\\
$^1$Department of Mathematics and Statistics, The University of Lahore,\\
1-KM Defence Road Lahore-54000, Pakistan.\\
$^2$Research Center of Astrophysics and Cosmology, Khazar University, \\
Baku, AZ1096, 41 Mehseti Street, Azerbaijan.}

\date{}
\maketitle

\begin{abstract}
In this paper, the notion of complexity factor and its implication
is extended to the framework of non-conserved Rastall theory of
gravity. First of all, the field equations governing a static
spherical geometry associated with the anisotropic fluid are
formulated. The mass function corresponding to the considered
geometry is defined in terms of both matter and geometric
quantities. The orthogonal decomposition of the Riemann tensor is
then performed through which a family of scalar quantities, known as
structure scalars, is obtained. Using the Herrera's recent
definition, one of the scalars among them is claimed as the
complexity factor, \emph{i.e.}, $\mathcal{Y}_{TF}$. Since there are
extra degrees of freedom in the gravitational equations, some
constraints are needed to make their solution possible to obtain. In
this regard, a well-known vanishing complexity condition is
introduced along with three different constraints which ultimately
lead to distinct stellar models. In order to check their physical
feasibility, a detailed graphical interpretation is provided using
multiple values of the Rastall parameter. It is concluded that the
obtained results in all three cases are consistent with those of
general relativity. Further, the Rastall theory provides more
suitable results in the case of model 2, indicating its superiority
over Einstein's gravity theory.
\end{abstract}

\section{Introduction}

It is widely accepted in cosmology that the universe is undergoing a
phase of an accelerated expansion. The latest observations, covering
type Ia supernovae \cite{1a,1b}, large-scale structures \cite{1c},
and fluctuations in the cosmic microwave background \cite{1d,1e},
have affirmed this development. In response to the challenge
regarding the cosmic accelerated expansion, astrophysicists have
turned to two major research directions: probing dark energy or
modifying the theory of general relativity (GR). The first notion
posits that a strange force, termed a ``dark energy'', is
responsible for such accelerated growth, however, its precise
properties are still unclear. To reveal the hidden properties of
dark energy, researchers have proposed numerous dynamic frameworks,
such as the cosmological constant, phantom models \cite{1f,1g},
tachyon fields \cite{1h}, the Chaplygin gas \cite{1i}, etc. Serving
as the ratio of pressure to density, the dark energy equation of
state is a key analytical tool for understanding the expansion of
the universe and its long-term evolution.

An alternative method suggesting the modification of GR, resulting
in the development of several extended frameworks. In 1972, Peter
Rastall, the pioneer behind the Rastall theory, proposed that the
energy-momentum tensor (EMT) with zero divergence in flat geometry
could have its nonzero value in curved or non-Euclidean spacetime
\cite{1aa}. In this modified gravity, unlike in GR, the Ricci scalar
is added to the field equations, introduced through a parameter
called the Rastall parameter. After the development of this theory,
the manner in which matter fields engage with gravity has shifted,
resulting in a strong interaction of the matter distribution with
the geometry. Following this approach, researchers have studied
multiple phenomena and confirmed that this theory competes at the
same level as other GR extensions obtained through modifying the
Einstein-Hilbert action \cite{2aa}-\cite{7aad}. An important trait
of the Rastall theory is that a solution to the modified governing
equations characterizing a perfect fluid based interior geometry
also holds in GR. Regarding black hole configurations, it is
intriguing that GR and Rastall theory lead to identical outcomes in
vacuum region.

The field equations within any gravitational framework serve to
entirely characterize a celestial structure. The challenge of
finding their solutions is intensifying, primarily due to the
sophisticated role of higher-order derivatives in the geometric
terms, particularly when matter-geometry coupling is taken into
account. One can find the solutions either via analytical methods or
numerical approaches, however, the latter strategy relying on
suitable initial/boundary conditions relevant to the given scenario.
In order to derive a solution, it is necessary to have some further
information regarding the local physical properties. As an example,
recent research on compact objects highlighted the possible
influence of various factors on the interior properties of celestial
bodies which are coupled with isotropic/anisotropic fluid setup
\cite{1}-\cite{3}. Other significant physical elements, such as
density fluctuations, shear, and dissipation flux have also a
notable impact on the stability/instability of the isotropic
pressure condition \cite{4}.

The anisotropic spherical object is governed by three independent
field equation, with five unknowns (a couple of metric potentials
along with a matter triplet). Achieving a solution requires the
addition of two further conditions, commonly expressed as an
empirical assumption regarding a physical parameter or an equation
of state that usually links physical quantities \cite{5,6}. The use
of a polytropic equation \cite{7,8} has been a popular method among
researchers for analyzing the physical properties of objects such as
white dwarfs \cite{9}. The scope of the study has been broadened to
examine anisotropic systems within GR \cite{10}-\cite{14da}. At the
same time, limitations on the metric potentials have been
established, with the Karmarkar condition as an example, where an
arbitrary function is chosen to derive a complete solution to the
given metric \cite{15}-\cite{20}. One more choice to be used is the
conformally flat geometry, where the Weyl tensor is rendered zero
\cite{21}. Such thorough investigations showed that the structural
development of a massive system can be framed in different ways,
each determined by its specific conditions.

The examination of complexity in heavily systems has surfaced as an
engaging and critical subject in current studies. This exploration
is fueled by the desire to understand the characteristics of compact
stars, in which gravitational forces control the particle
interactions. The range of these systems includes various types of
entities, from star clusters and galaxies in the astronomical
context to extensive structures in the universe, like the
distribution of matter on large scales. The complexity seen in
celestial systems is driven by multiple factors such as the
non-linear dynamics in which even small changes in influential
parameters can cause major differences in the evolution of structure
and unexpected consequences. The concept of complexity stresses the
diverse components within the fluid distribution, which collectively
contribute to its complex behavior. The elements involved might be
the inhomogeneous density, anisotropic pressure, energy flux, and
other related quantities. Scholars have made attempts to define
complexity in a way that can be applied across diverse areas of
science. In its early stages, the complexity was conceptualized
based on the information content and entropy within the system under
scrutiny \cite{22}-\cite{24}. The application of this definition,
when applied on two contrasting systems, namely a perfect crystal
and an ideal gas, highlighted its limitations, as their
characteristics are inherently contradictory, though none of these
systems exhibits complexity.

A very recent and widely recognized definition of complexity was put
forwarded by Herrera \cite{25}. In his analysis, he associated the
concept of complexity with physical parameters, including energy
density variations and anisotropy in the principal pressure.
Applying Bel's idea of decomposing the Riemann tensor into its
orthogonal components \cite{26,27}, Herrera found a class of
structure scalars. He highlighted that the two key elements
discussed earlier are present in a single scalar function,
$\mathcal{Y}_{TF}$, which he named the complexity factor. This
notion was initial relied on the assumption that complexity inside a
stellar object declines in two situations: when the interior
spacetime is associated with homogeneous/isotropic properties, or
when the pressure anisotropy and energy density inhomogeneity cancel
each other's impact. The approach has been widened to investigate
the evolutionary patterns for certain objects having non-static and
axial geometries using the same complexity factor initially claimed
only for a static spherical structure \cite{28,28a}.

Sharif and Naseer applied this concept to discuss charged spherical
systems possessing a static interior within GR as well as a modified
minimally coupled framework, revealing that $\mathcal{Y}_{TF}$ acts
as a complexity factor even in both these scenarios \cite{30,31}. In
their study, Abbas and Nazar \cite{32} analyzed new solutions within
the $f(\mathbf{R})$ framework, examining how modified terms affect
the complexity and the mass function of the spherical system. Their
study on compact stars incorporated a complexity-free condition and
the Krori-Barua ansatz, within a minimally interacted model and
acceptable results have been found \cite{33}. Manzoor and Shahid
\cite{34} looked into the evolutionary changes of the compact object
$4U1820-30$ by utilizing the Starobinsky model. It was confirmed
that, at certain parameter values, the dark matter density exceeds
the conventional matter density by a ratio greater than $1$. Some
interesting works in this regard can be found in
\cite{34a}-\cite{34d}.

This study expands on earlier anisotropic models \cite{40} and the
notion of complexity within the Rastall gravity. The layout of this
article is structured in the following. The basics of this
non-conserved theory and the anisotropic EMT are presented in
section \textbf{2}. Further, the modified field equations and the
mass function are developed under a static spherical metric. The
matching conditions are also evaluated at the boundary by connecting
the interior and exterior geometries. Section \textbf{3} discusses
the structure scalars, and $\mathcal{Y}_{TF}$ is pointed as the
complexity factor for the current analysis. In section \textbf{4}, a
concise overview of the necessary conditions is presented which must
be fulfilled for a physically consistent stellar model. Section
\textbf{5} presents three distinct models, each of them is
interpreted graphically under certain choices of the Rastall
parameter. The concluding section provides a summary of the obtained
results along with their comparison with other gravity theories.

\section{Fundamentals of Rastall Theory and Spherical Spacetime}

By deviating from the fundamental idea that the EMT's divergence
vanishes in flat spacetime, rather this does not disappear in curved
geometry, the Rastall theory has been introduced \cite{1aa}. The
field equations corresponding to this recently established theory
are
\begin{equation}\label{g1}
G_{\varphi\eta}\equiv
R_{\varphi\eta}-\frac{1}{2}Rg_{\varphi\eta}=\kappa
\big(T_{\varphi\eta}-\omega Rg_{\varphi\eta}\big),
\end{equation}
where
\begin{itemize}
\item $G_{\varphi\eta}$ is used to denote the Einstein tensor,
\item $\kappa$ signifies the coupling constant,
\item The Rastall parameter, represented by $\omega$,
results in a distinction from Einstein's field equations.
\end{itemize}
A key observation is that these field equations match the
non-conservation expressed by
\begin{equation}\label{g2}
\nabla^\varphi T_{\varphi\eta}=\omega g_{\varphi\eta}\nabla^\varphi
R,
\end{equation}
when $\omega = 0$ is substituted, it simplifies to GR. The non-zero
divergence in Eq.\eqref{g2} must be emphasized, as it is vital to
formulating the fluid-geometry coupling. If we redefine the factor
on right side of Eq.\eqref{g1}, it may be written as
\begin{equation}\label{g3}
\tilde{T}_{\varphi\eta}=T_{\varphi\eta}-\omega Rg_{\varphi\eta},
\end{equation}
thus, make Eq.\eqref{g1} in the following manner
\begin{eqnarray}\label{g4}
G_{\varphi\eta}=\kappa \tilde{T}_{\varphi\eta}.
\end{eqnarray}
Notably, if we follow the above field equations, they give rise to
the standard conservation equation $\nabla^\varphi
\tilde{T}_{\varphi\eta} = 0$. It is well-documented that such an
approach can be easily accomplished in other modified theories,
where generalized functions of geometric and matter terms replace
the Ricci scalar in the Einstein-Hilbert action.

At this stage of the analysis, the goal is to modify the Rastall
field equations. The trace of Eq.\eqref{g1} produces
\begin{eqnarray}\label{g4a}
R=\frac{\kappa T}{4\omega\kappa-1}.
\end{eqnarray}
Substituting this into Eq.\eqref{g3}, the effective EMT takes the
form of
\begin{equation}\label{g4b}
\tilde{T}_{\varphi\eta}=T_{\varphi\eta}-\frac{\alpha T}{4\alpha-1}
g_{\varphi\eta},
\end{equation}
where $\alpha=\omega\kappa$. We set $\kappa = 1$, leading to the
condition $\alpha=\omega$. It should be emphasized that realistic
configurations can only be achieved when $\alpha \neq \frac{1}{4}$.

The study of Rastall gravity is undoubtedly nuanced and involves
various perspectives. The contrasting opinions of Visser \cite{8aa}
and Golovnev \cite{baa}, who argued that Rastall gravity actually
represents a redefinition of the EMT of GR, are met with rebuttals
from Darabi \emph{et al.} \cite{caa}, who suggested that this theory
involves a non-minimal coupling, marking a distinction from GR. In
Visser's view \cite{8aa}, the Rastall EMT $\tilde{T}_{\varphi\eta}$
allows for the reconstruction of the physical quantity
$T_{\varphi\eta}$, and the process works in reverse as well. We take
a different view, as if the aim is to demonstrate that Rastall
theory is equivalent to GR, this can similarly be done for other
matter-geometry coupled theories, including $f(R,T)$ gravity, which
emerged from the modification of the Einstein-Hilbert action. As a
result, it cannot be considered a valid point to assert that GR and
Rastall theory are identical.

The role of anisotropy in the interior of compact stellar objects is
a significant area of astrophysical investigation, as it affects
both their structural integrity and dynamic evolution. The term
anisotropy describes the difference in the pressure distribution
within a star, with the radial pressure differing from the
tangential one. Within high-density environments like neutron stars,
anisotropy is essential, as powerful gravitational and magnetic
fields result in significant departures from isotropy. The
investigation of anisotropic models has progressed notably since the
early 20$^{\text{th}}$ century, with essential contributions
focusing on the inclusion of anisotropic pressure in stellar
studies. As an illustration, multiple studies have laid the
groundwork for comprehending the development of anisotropic
pressures, driven by factors such as stellar rotation, magnetic
fields, phase transitions, and other contributing elements
\cite{37n}-\cite{37rb}. The EMT associated with this distribution is
represented as
\begin{equation}\label{g5}
T_{\varphi\eta}=(\rho+p_t)v_{\varphi}v_{\eta}+p_t
g_{\varphi\eta}+\left(p_r-p_t\right)u_\varphi u_\eta,
\end{equation}
with $\rho$, $p_t$ and $p_r$ being the energy density, tangential
and radial pressures, respectively. Also, $v_{\varphi}$ symbolizes
the four-velocity and $u_{\varphi}$ being the four-vector.

For the purpose of investigating complexity for compact fluid setup,
the interior spacetime is considered to follow a spherical geometry,
represented as
\begin{equation}\label{g6}
ds^2=-e^{\delta_1} dt^2+e^{\delta_2}
dr^2+r^2\big(d\theta^2+\sin^2\theta d\phi^2\big),
\end{equation}
where $\delta_1 = \delta_1(r)$ and $\delta_2 = \delta_2(r)$,
indicating the static nature of the considered geometry. The values
of the four-quantities presented in Eq.\eqref{g5} are computed for
the above metric as
\begin{equation}\label{g7}
u^\varphi=\delta^\varphi_1e^{\frac{-\delta_2}{2}}, \quad
v^\varphi=\delta^\varphi_0e^{\frac{-\delta_1}{2}},
\end{equation}
where the relations $u^\varphi v_{\varphi} = 0$, $u^\varphi
u_{\varphi} = 1$, and $v^\varphi v_{\varphi} = -1$ hold true. The
non-null components of Rastall field equations are obtained using
Eqs.\eqref{g4}-\eqref{g6} as
\begin{align}\label{g8}
&e^{-\delta_2}\left(\frac{\delta_2'}{r}-\frac{1}{r^2}\right)
+\frac{1}{r^2}=\rho-\frac{\alpha}{4\alpha-1}\left(\rho-p_r-2p_t\right),\\\label{g9}
&e^{-\delta_2}\left(\frac{1}{r^2}+\frac{\delta_1'}{r}\right)
-\frac{1}{r^2}=p_r+\frac{\alpha}{4\alpha-1}\left(\rho-p_r-2p_t\right),
\\\label{g10}
&\frac{e^{-\delta_2}}{4}\left[\delta_1'^2-\delta_1'\delta_2'+2\delta_1''-\frac{2\delta_2'}{r}+\frac{2\delta_1'}{r}\right]
=p_t+\frac{\alpha}{4\alpha-1}\left(\rho-p_r-2p_t\right),
\end{align}
in which the terms combined with $\frac{\alpha}{4\alpha-1}$ refer to
the Rastall's corrections and $'=\frac{\partial}{\partial r}$. The
explicit expressions of the fluid triplet are expressed by
\begin{align}\label{g8a}
\rho&=\frac{e^{-\delta_2}}{2r^2}\big[r \delta_2 ' \big(2-4 \alpha
-\alpha  r \delta_1 '\big)-2 (2 \alpha -1) \big(e^{\delta_2
}-1\big)+\alpha  r \big\{2 r \delta_1 ''+\delta_1 ' \big(r \delta_1
'+4\big)\big\}\big],\\\label{g9a}
p_r&=\frac{e^{-\delta_2}}{2r^2}\big[r \big\{\alpha  \big(r \delta_1
'+4\big) \big(\delta_2 '-\delta_1 '\big)-2 \alpha  r \delta_1 ''+2
\delta_1 '\big\}+2 (2 \alpha -1) \big(e^{\delta_2
}-1\big)\big],\\\label{g10a} p_t&=\frac{e^{-\delta_2}}{4r^2}\big[r
\big\{\big(8 \alpha +(2 \alpha -1) r \delta_1 '-2\big) \big(\delta_2
'-\delta_1 '\big)+2 (1-2 \alpha ) r \delta_1 ''\big\}+8 \alpha
\big(e^{\delta_2 }-1\big)\big].
\end{align}

Expanding the non-conservation relation $\nabla_\varphi
T^{\varphi\eta} \neq 0$ with the metric \eqref{g6}, we derive the
following expression, valid within the Jordan frame (for a detailed
explanation, consult \cite{40a}) as
\begin{align}\label{g12}
\frac{dp_r}{dr}+\frac{\delta_1'}{2}\left(\rho+p_r\right)+\frac{2\Pi}{r}
=\frac{\alpha}{4\alpha-1}\big(p_r'+2p_t'-\rho'\big).
\end{align}
Here, $\Pi=p_r-p_t$ denotes the pressure anisotropy, and
Eq.\eqref{g12} is recognized as the generalized
Tolman-Oppenheimer-Volkoff equation for matter distributions coupled
with the anisotropic fluid \cite{40b}. The above equation is very
important tool for probing the evolutionary modifications in a
self-gravitating system. It can be noted that the Rastall equations
of motion \eqref{g8}-\eqref{g10} now contain a high degrees of
freedom, implying the presence of several unknowns, such as
$(\delta_1, \delta_2, \rho, p_r, p_t)$. Consequently, applying
simultaneous constraints on this system is required to find a unique
non-singular solution.

For a static sphere, the mass function is formulated based on the
Misner-Sharp definition and can be written as follows
\begin{align}\label{g12a}
m(r)=\frac{r}{2}\big(1-e^{-\delta_2}\big).
\end{align}
This function can alternatively be expressed in the form of fluid
terms as
\begin{align}\label{g12b}
m(r)&=\frac{1}{2}\int_0^r\left\{\rho-\frac{\alpha}{4\alpha-1}\left(\rho-p_r-2p_t\right)\right\}\bar{r}^2d\bar{r}.
\end{align}
The calculation of $\delta_1'$ (present in Eq.\eqref{g12}) in terms
of the matter sector and mass function can be carried out by
combining \eqref{g9} and \eqref{g12a}. This is demonstrated as
\begin{align}\label{g12c}
\delta_1'=\frac{1}{r\big(r-2m\big)}
\bigg[\bigg\{p_r+\frac{\alpha}{4\alpha-1}\left(\rho-p_r-2p_t\right)\bigg\}r^3+2m\bigg].
\end{align}
By substituting this value into Eq.\eqref{g12}, we obtain the
following result
\begin{align}\nonumber
&\frac{dp_r}{dr}+\frac{\rho+p_r}{2r\big(r-2m\big)}\bigg[\bigg\{p_r+\frac{\alpha}{4\alpha-1}
\left(\rho-p_r-2p_t\right)\bigg\}r^3+2m\bigg]\\\label{g12d}
&+\frac{2\Pi}{r}+\frac{\alpha}{4\alpha-1}\big(\rho'-p_r'-2p_t'\big)=0.
\end{align}

To connect the interior and exterior metrics smoothly at the
boundary, the junction conditions are employed, leading to a
comprehensive solution that helps to understanding the evolution of
the structure more comprehensively. The outer geometry is described
by considering the Schwarzschild metric with total mass $\texttt{M}$
given below
\begin{equation}\label{g15}
ds^2=-\bigg(1-\frac{2\texttt{M}}{r}\bigg)dt^2+\bigg(1-\frac{2\texttt{M}}{r}\bigg)^{-1}dr^2
+r^2\big(d\theta^2+\sin^2\theta d\phi^2\big).
\end{equation}
The smooth matching of these metrics is contingent upon the
following relations at $r = r_\Sigma = \texttt{R}$. The two
fundamental forms of junction conditions give rise to these, as
shown by
\begin{align}\label{g16}
e^{\delta_1}~{_=^\Sigma}~1-\frac{2\texttt{M}}{\texttt{R}}~{_=^\Sigma}~e^{-\delta_2},
\quad p_r~{_=^\Sigma}~0.
\end{align}
The two equations on left side match the radial/time components of
both the exterior and interior spacetimes at the spherical
interface, and the remaining one enforces the requirement that the
radial pressure must be zero at $r=\texttt{R}$.

\section{Structure Scalars and Adoption of Complexity Factor}

The idea of complexity in self-gravitating systems has turned into a
primary focus of astrophysical studies. In the literature,
complexity has been defined in many ways, with one definition
suggesting that a homogenous/isotropic structure holds a complexity
of zero value. Measuring the complexity clearly captures how
irregularities in energy density correlate with the anisotropy in
pressure. Herrera \cite{25} defined this concept using some scalar
factors by decomposing the curvature tensor \cite{26,27} and
selecting one of its components as the complexity factor. We shall
briefly present an overview of the method for determining the
complexity factor in the following, underlining its relevance to the
physical parameters stated above. The equation that illustrates the
splitting of $R^{\zeta\varphi}_{\beta\vartheta}$ is expressed by
\begin{equation}\label{g23}
R^{\zeta\varphi}_{\beta\vartheta}=C^{\zeta\varphi}_{\beta\vartheta}+2
T^{[\zeta}_{[\beta}\delta^{\varphi]}_{\vartheta]}+
T\left(\frac{1}{3}\delta^{\zeta}_{[\beta}\delta^{\varphi}_{\vartheta]}
-\delta^{[\zeta}_{[\beta}\delta^{\varphi]}_{\vartheta]}\right),
\end{equation}
where $C^{\zeta\varphi}_{\beta\vartheta}$ is used to denote the Weyl
tensor. It must be noted that the procedure of the orthogonal
splitting of the Riemann curvature tensor is indeed same, however,
the results produced for structure scalars are different in modified
theories as the energy density, radial and tangential pressures
involve correction terms which directly affect the profile of the
scalar quantities, and ultimately making a self-gravitating object
more or less complex. Furthermore, both tensors are formulated in
the following manner
\begin{eqnarray}\label{g24}
\mathcal{Y}_{\zeta\beta}&=&R_{\zeta\varphi\beta\vartheta}v^{\varphi}v^{\vartheta},\\\label{g25}
\mathcal{X}_{\zeta\beta}&=&^{\ast}R^{\ast}_{\zeta\varphi\beta\vartheta}v^{\varphi}v^{\vartheta}
=\frac{1}{2}\eta^{\omega\sigma}_{\zeta\varphi}R^{\ast}_{\omega\sigma\beta\vartheta}
v^{\varphi}v^{\vartheta},
\end{eqnarray}
where $\eta^{\omega\sigma}_{\zeta\varphi}$ refers to the Levi-Civita
symbol and $R^{\ast}_{\zeta\varphi\beta\vartheta} = \frac{1}{2}
\eta_{\omega\sigma\beta\vartheta} R^{\omega\sigma}_{\zeta\varphi}$.
Equations \eqref{g24} and \eqref{g25} may also be represented in a
different fashion as
\begin{eqnarray}\label{g26}
\mathcal{Y}_{\zeta\beta}&=&\frac{1}{3}\big\{h_{\zeta\beta}\mathcal{Y}_{T}+\big(3v_{\zeta}v_{\beta}
-h_{\zeta\beta}\big)\mathcal{Y}_{TF}\big\},\\\label{g27}
\mathcal{X}_{\zeta\beta}&=&\frac{1}{3}\big\{h_{\zeta\beta}\mathcal{X}_{T}+\big(3v_{\zeta}v_{\beta}
-h_{\zeta\beta}\big)\mathcal{X}_{TF}\big\}.
\end{eqnarray}
We define the projection tensor as $h_{\zeta\beta} = u_{\zeta}
u_{\beta} + g_{\zeta\beta}$. Performing some simple but detailed
calculations (which are not presented here) based on
Eqs.\eqref{g23}-\eqref{g27}, we arrive at four scalars expressed by
\begin{eqnarray}\label{g28}
&&\mathcal{X}_{T}=\rho,\\\label{g28a}
&&\mathcal{X}_{TF}=-\mathbf{E}-\frac{\Pi}{2},\\\label{g28b}
&&\mathcal{Y}_{T}=\frac{1}{2}\big(\rho+3p_r-2\Pi\big),\\\label{g28c}
&&\mathcal{Y}_{TF}=\mathbf{E}-\frac{\Pi}{2}.
\end{eqnarray}
In the above scalars, the electric component of the Weyl tensor,
represented by $\mathbf{E}$, is the only undefined physical factor.
Its value is presented as follows
\begin{equation}\label{g29}
\mathbf{E}=\frac{e^{-\delta_2}}{4}\left[\delta_1''+\frac{\delta_1'^2-\delta_2'\delta_1'}{2}
-\frac{\delta_1'-\delta_2'}{r}+\frac{2(1-e^{\delta_2})}{r^2}\right].
\end{equation}
Our observations suggest that these scalar functions are closely
related to the physical features governing celestial interiors which
are coupled with both uniform/non-uniform density distributions as
well as anisotropic properties. By examining
Eqs.\eqref{g28}-\eqref{g28c}, the evolution of a self-gravitating
structure can be better understood, as explained in the subsequent
lines
\begin{itemize}
\item The expression $\mathcal{X}_{T}$ explains the effect of
homogeneous energy density on the fluid's distribution,
\item $\mathcal{X}_{TF}$ specifies the degree of energy density
inhomogeneity,
\item $\mathcal{Y}_{T}$ regulates the local anisotropy,
\item $\mathcal{Y}_{TF}$ acts in the capacity of both $ \mathcal{X}_{TF} $ and $ \mathcal{Y}_{T} $.
\end{itemize}

In order to emphasize the complexity factor among them, we need to
perform certain manipulations. To address this, the mass function,
matter variables, and the Weyl scalar can be tied together as shown
below
\begin{align}\nonumber
m&=\frac{r^3}{6}\big(\rho-p_r+p_t\big)
-\frac{r^3\mathbf{E}}{3}+\frac{e^{-\delta_2}r}{3}\big(e^{\delta_2}-1\big)-\frac{e^{-\delta_2}r}{24}\big[r
\big\{\big(\delta_1'-\delta_2'\big)\\\label{g29a} &\times \big(2 (4
\alpha -1)+2 \alpha r \delta_1 '+2\big)+4 \alpha  r \delta_1
''\big\}-8 (\alpha -1) \big(e^{\delta_2 }-1\big)\big].
\end{align}
As a result of combining this with Eq.\eqref{g12b}, we obtain the
following
\begin{align}\nonumber
\mathbf{E}&=\frac{e^{-\delta_2}}{4 r^3}\bigg[\alpha  r \big\{r
\big(\big(r \delta_1 '+4\big) \big(\delta_2 '-\delta_1'\big)-2 r
\delta_1 ''\big)-4\big\}\\\label{g29b} &-2 e^{\delta_2} \big\{r^3
(\Pi -\rho )-2 \alpha
r\big\}\bigg]-\frac{3}{2r^3}\int_0^r\tilde{\rho}\bar{r}^2d\bar{r},
\end{align}
with
$\tilde{\rho}=\rho-\frac{\alpha}{4\alpha-1}\left(\rho-p_r-2p_t\right)$.
When this is substituted into Eq.\eqref{g28c}, it becomes clear that
this factor alone encapsulates all the necessary parameters to
define the complexity, expressed as follows
\begin{align}\nonumber
\mathcal{Y}_{TF}&=\frac{e^{-\delta_2}}{4 r^3}\bigg[\alpha  r \big\{r
\big(\big(r \delta_1 '+4\big) \big(\delta_2 '-\delta_1'\big)-2 r
\delta_1 ''\big)-4\big\}\\\label{g30} &-2 e^{\delta_2} \big\{r^3
(2\Pi -\rho )-2 \alpha
r\big\}\bigg]-\frac{3}{2r^3}\int_0^r\tilde{\rho}\bar{r}^2d\bar{r}.
\end{align}
One must observe that this factor becomes zero when the matter
content is assumed to be isotropic and homogeneous. Moreover, a
recently proposed formula for the Tolman mass facilitates the
estimation of a structure's total mass-energy, although it does not
guarantee its localization \cite{aa}.

It is important to highlight that a structure with zero complexity
is not determined solely by homogeneous/isotropic setup. This can be
achieved through the condition $\mathcal{Y}_{TF} = 0$. Using this
along with Eq.\eqref{g30}, the following expression is obtained,
linking governing fluid parameters with each other as
\begin{equation}\label{g34}
\Pi=\frac{e^{-\delta_2}}{2r^3}\bigg[\alpha r \big\{r \big(\big(r
\delta_1 '+4\big) \big(\delta_2 '-\delta_1'\big)-2 r \delta_1
''\big)-4\big\}+2re^{\delta_2}\big(\rho
r^2+2\alpha\big)-4e^{\delta_2}\int_0^r\tilde{\rho}\bar{r}^2d\bar{r}\bigg],
\end{equation}
implying that a class of solutions exists that satisfy this
condition \cite{40}. Given that this condition corresponds to a
non-local equation of state \cite{ai}, this method is especially
effective in solving the governing equations \eqref{g2}. The ability
to satisfy the relevant constraint in a realistic spherically
symmetric system is an exciting area for future research. The goal
is to investigate this aspect by analyzing situations where
anisotropic EMT naturally emerges, such as in environments with
exotic matter distributions or in the vicinity of extremely dense
objects like white dwarfs and black holes. Analyzing these scenarios
can offer valuable insights into the relevance and effects of this
condition \eqref{g34} in actual astrophysical interiors. Hence,
according to \cite{22}, ignoring this element may result in entirely
distinct systems.

\section{Substantial Requirements for Self-gravitating Models to Exist}

Several approaches have been explored by various researchers to deal
with the field equations governing physically significant celestial
bodies. When these solutions do not meet the acceptable criteria,
they are deemed irrelevant for modeling actual compact stars. A
variety of conditions in this regard have been proposed and
maintained by different researchers \cite{ab,ac}. The conditions
outlined below provide further details in this regard.
\begin{itemize}
\item For a self-gravitating fluid system, the radial and temporal
metric functions should remain finite, devoid of singularities, and
always positive.

\item The maximum value of matter variables, including energy density and pressure,
should occur at the center ($r = 0$), revealing a steady and
positive progression throughout the whole range. Their first
(second) derivatives must be zero (negative) at $r = 0$ and show a
decreasing behavior as they extend toward the boundary.

\item \textcolor{blue} {The manner in which particles are arranged in a compact body
determines their proximity to each other. Through this, the compact
nature of the system can be defined. Another definition is the
mass-to-radius proportion, which must be beneath the outlined limit
for the case of a static sphere in GR, as mentioned in
\cite{ad,adad}
\begin{align}\label{g49}
\frac{2\texttt{M}}{\texttt{R}} < \frac{8}{9}.
\end{align}
However, when extending to modified theories of gravity or
introducing pressure anisotropy in the fluid distribution, this
limit is subject to modifications due to changes in gravitational
dynamics and internal pressure profiles \cite{adada}-\cite{adade}.
Anisotropy, characterized by unequal radial and tangential
pressures, significantly influences the stability and compactness of
stellar structures. It is well-established in the literature that
anisotropy can alter the Buchdahl limit by introducing additional
forces that either counteract or enhance gravitational collapse.
Specifically, positive anisotropy generates outward-directed forces,
which can stabilize compact structures and potentially raise the
upper bound of compactness. On the other hand, negative anisotropy
contributes inward-directed forces, which may lower the compactness
threshold. Moreover, the non-minimal coupling in Rastall theory
alters gravitational dynamics, impacting stellar properties such as
density, pressure profiles, and compactness. The current study is
intended to numerically solving highly non-linear differential
equations arising from Rastall gravity. Consequently, the explicit
analytical expressions for the modified Buchdahl limit cannot be
derived. The numerically computed values of this factor under
certain parametric choices (\emph{i.e.}, $\alpha = 0,~0.1,~0.2$) are
provided in Table \textbf{1}. It must be highlighted that the
Buchdahl limit is found to be slightly lesser in Rastall theory as
compared to GR.}
\begin{table}[H]
\scriptsize \centering \caption{\textcolor{blue} {Numerical values
of Buchdahl limit for different values of Rastall parameter.}}
\label{Table1} \vspace{+0.07in} \setlength{\tabcolsep}{1.4em}
\textcolor{blue} {\begin{tabular}{cccccc}
\hline\hline \textbf{Values of $\alpha$} & 0 & 0.1 & 0.2 & 0.3 & 0.4
\\\hline
\textbf{Buchdahl limit $(2\texttt{M}/\texttt{R})$} & $\approx$ 0.889
& $\approx$ 0.772 & $\approx$ 0.714 & $\approx$ 0.652 & $\approx$
0.636
\\
\hline\hline
\end{tabular}}
\end{table}

\item The presence of ordinary matter in a celestial body's interior
is assured by satisfying certain conditions. These are identified as
energy conditions, which are combinations of the governing fluid
parameters within the corresponding EMT. For the system under
consideration, these conditions are
\begin{equation}
\left.
\begin{aligned}\label{g50}
&\rho+p_r \geq 0, \quad \rho+p_t \geq 0, \\
&\rho-p_r \geq 0, \quad \rho-p_t \geq 0,\\
&\rho \geq 0, \quad \rho+p_r+2p_t \geq 0.
\end{aligned}
\right\}
\end{equation}
The dominant energy constraints (\emph{i.e.}, $\rho - p_{r} \geq 0$
and $\rho - p_{t} \geq 0$) are key conditions, affirming that the
energy density must always be larger than both the tangential/radial
pressures.

\item In the interior fluid model, the gravitational redshift is
formulated as $z = e^{-\delta_1 / 2} - 1$. As the factor is solely
dependent on the $g_{tt}$ metric potential, it is imperative that it
decreases with increasing radius. For the model to be valid, its
value must remain below $5.211$ at $r=\texttt{R}$ \cite{ad}.

\item Investigating the stability of systems with small deviations
from hydrostatic equilibrium has attracted significant interest in
current research. The phenomenon of cracking, firstly introduced by
Herrera \emph{et al.} \cite{ae,af}, occurs when the total radial
force switches sign at a particular location because of some
external factors. For cracking to be prevented, the following
inequality must be true
\begin{align}\label{g51}
0 \leq v_{r}^{2}-v_{t}^{2} \leq 1,
\end{align}
In this case, $v_{t}^{2} = \frac{dp_{t}}{d\rho}$ refers to the
tangential, while $v_{r}^{2} = \frac{dp_{r}}{d\rho}$ signifies the
radial sound speed.
\end{itemize}

\section{Development of a Class of New Stellar Models}

Several models have been proposed by scholars to solve the field
equations. To illustrate, Herrera \cite{25} found solutions by
utilizing two different limitations (such as the Gokhroo-Mehra
ansatz and the polytropic equation). By applying distinct conditions
in the subsequent subsections, we derive three solutions and assess
their physical properties using graphical interpretation for
different values of the parameters involved in this study.

\subsection{Radial Pressure and Complexity Factor to be Null}

To solve the field equations \eqref{g8a}-\eqref{g10a}, which involve
five unknowns ($\rho$, $p_r$, $p_t$, $\delta_1$, $\delta_2$),
additional constraints are necessary. To reduce the complexity, we
implement the assumptions $\mathcal{Y}_{TF} = 0$ and $p_r = 0$,
yielding a spherical solution that contains only the tangential
component of pressure, a concept in line with Florides \cite{ah}.
The criteria mentioned earlier gives the following differential
equation when merged with \eqref{g9a} as
\begin{align}\label{g52}
& r \big\{\alpha\big(r \delta_1 '+4\big) \big(\delta_2 '-\delta_1
'\big)-2 \alpha  r \delta_1 ''+2 \delta_1 '\big\}+2 (2 \alpha -1)
\big(e^{\delta_2}-1\big)=0.
\end{align}

On the other hand, combining Eqs.\eqref{g8a}, \eqref{g10a}, and
\eqref{g34}, the complexity-free constraint takes the form
\begin{align}\nonumber
&r \big[\delta_1 ' \big\{\alpha  r^2 \big(\delta_2 ''-2 \delta_1
''\big)+8 \alpha (r-1)-2 r\big\}+2 r \big\{(2 \alpha -1) \delta_2
''-\alpha r \delta_1 ^{(3)}\\\nonumber &+(2 \alpha  r-r-5 \alpha)
\delta_1 ''\big\}-r \delta_2 '^2 \big(4 \alpha +\alpha  r \delta_1
'-2\big)+\delta_2 ' \big\{r \big(3 \alpha r \delta_1 ''+\delta_1 '
\\\nonumber &\times \big(7
\alpha +\alpha  r \delta_1 '-2 \alpha r+r\big)\big)+2 (6 \alpha -4
\alpha r+r-3)\big\}+r \{(2 \alpha -1) r\\\label{g53} &-3 \alpha \}
\delta_1 '^2\big]-2 (\alpha (4 r-2)+1) \big(e^{\delta_2 }-1\big)=0.
\end{align}
The presence of higher-order curvature terms in the adjusted
formulation leads to the previous two equations being fourth-order
in the spacetime potentials, which implies that an exact solution
cannot be achieved. In order to find the values of $g_{rr}$ and
$g_{tt}$, we numerically integrate Eqs. \eqref{g52} and \eqref{g53}
with carefully chosen initial conditions. As shown in Figure
\textbf{1}, the profiles of $e^{\delta_1}$ and $e^{-\delta_2}$
exhibit a positive and non-singular trend at all points. At the
central point, the values $e^{\delta_1(0)} = c_1$ (with $c_1 > 0$)
and $e^{-\delta_2(0)} = 1$ are consistent with what was expected.
The star's radius or outermost limit is defined as the point where
these two functions intersect. For this model, we retrieve these
values and the compactness factor as \textcolor{blue}
{\begin{itemize}
\item At $\alpha = 0$, we get $\texttt{R} = 0.12$, and the compactness value is
$$e^{\delta_1(0.12)}=e^{-\delta_2(0.12)}\approx0.28=1-\frac{2\texttt{M}}{\texttt{R}}
\quad \Rightarrow \quad \frac{2\texttt{M}}{\texttt{R}} \approx0.72 <
0.889.$$
\item At $\alpha = 0.1$, we get $\texttt{R} = 0.14$, and the compactness is
$$e^{\delta_1(0.14)}=e^{-\delta_2(0.14)}\approx0.36=1-\frac{2\texttt{M}}{\texttt{R}}
\quad \Rightarrow \quad \frac{2\texttt{M}}{\texttt{R}} \approx0.64 <
0.772.$$
\item Setting $\alpha = 0.2$ results in $\texttt{R} = 0.19$, and the compactness takes the value
$$e^{\delta_1(0.19)}=e^{-\delta_2(0.19)}\approx0.82=1-\frac{2\texttt{M}}{\texttt{R}}
\quad \Rightarrow \quad \frac{2\texttt{M}}{\texttt{R}} \approx0.18 <
0.714.$$
\end{itemize}}
\begin{figure}\center
\epsfig{file=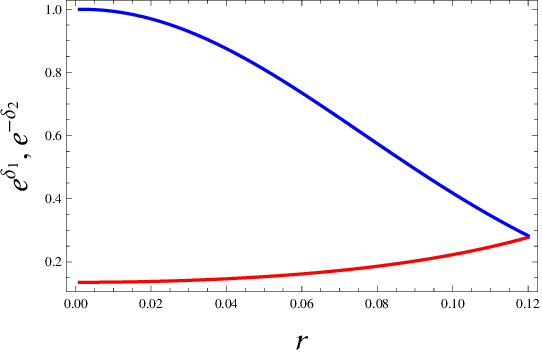,width=0.4\linewidth}\epsfig{file=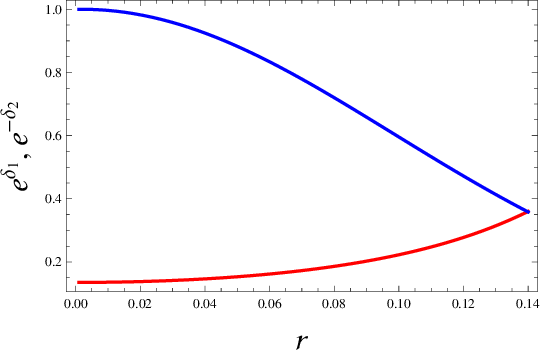,width=0.4\linewidth}
\epsfig{file=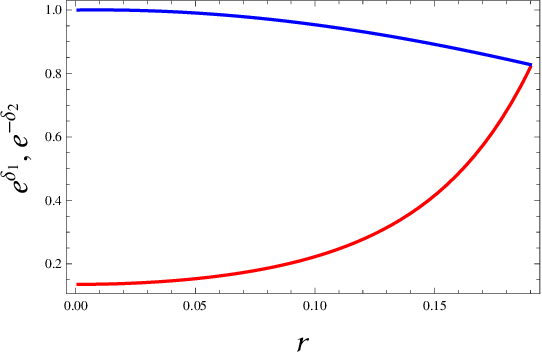,width=0.4\linewidth}
\caption{Metric potentials $e^{\delta_1}$
(\textcolor{red}{\textbf{\small $\bigstar$}}) and $e^{-\delta_2}$
(\textcolor{blue}{\textbf{\small $\bigstar$}}) for $\alpha=0$ (upper
left), $0.1$ (upper right) and $0.2$ (lower) analogous to model I.}
\end{figure}

As shown in Figure \textbf{2}, the energy density follows a trend
with a peak at $r = 0$, gradually diminishing towards the outer
surface. In addition, the interior structure affected by Rastall
corrections shows a decrease in density relative to the GR case
\cite{40}. The absence of radial pressure, as seen in Florides'
solution, it signifies that stability can be preserved until the
tangential pressure increases as the radius expands. The behavior of
$p_t$ is presented in the right plot, and it conforms to the
expected profile. The anisotropy shows a contrasting behavior to the
tangential pressure, as represented by $\Pi = -p_t$ (lower plot).

The energy conditions, presented in Figure \textbf{3}, display a
favorable trend, validating the first solution. In the left plot of
Figure \textbf{4}, the gravitational redshift is depicted, showing a
decrease as $r$ increases. Its values on the boundary are calculated
as $z(0.12) \approx 0.882$, $z(0.14) \approx 0.611$, and $z(0.19)
\approx 0.084$, corresponding to $\alpha = 0$, $0.1$, and $0.2$.
These values are considerably smaller than the upper limit observed
by researchers, \emph{i.e.}, $z({\texttt{R}}) = 5.211$. The
stability criterion is examined in the right plot, showing that the
model remains free from cracking, thus preserving its stability.
\begin{figure}\center
\epsfig{file=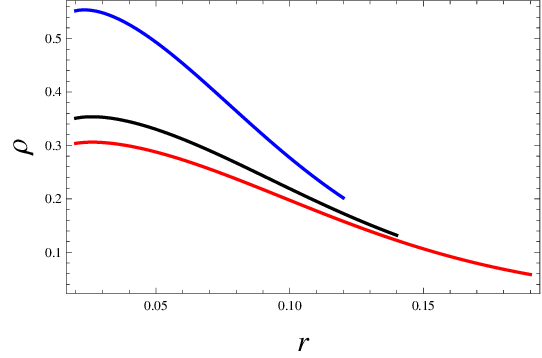,width=0.4\linewidth}\epsfig{file=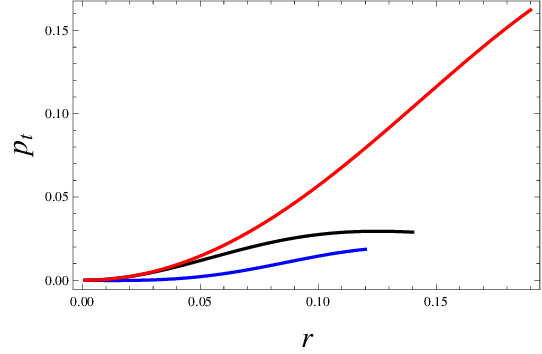,width=0.4\linewidth}
\epsfig{file=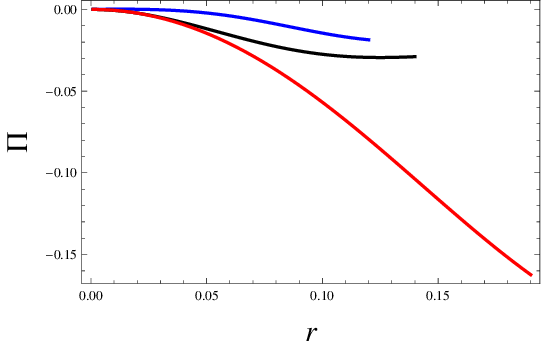,width=0.4\linewidth} \caption{Governing
parameters for $\alpha=0$ (\textcolor{blue}{\textbf{\small
$\bigstar$}}), $0.1$ (\textcolor{black}{\textbf{\small $\bigstar$}})
and $0.2$ (\textcolor{red}{\textbf{\small $\bigstar$}})
corresponding to model I.}
\end{figure}
\begin{figure}\center
\epsfig{file=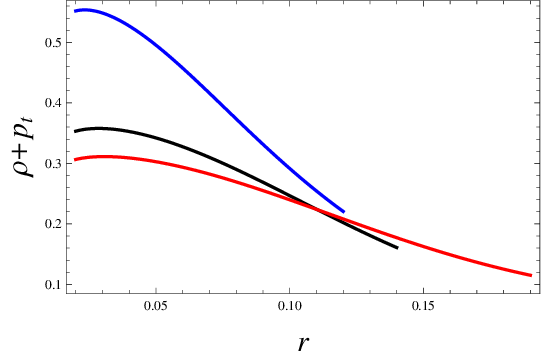,width=0.4\linewidth}\epsfig{file=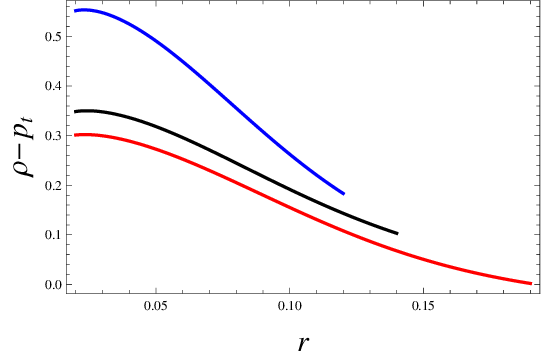,width=0.4\linewidth}
\epsfig{file=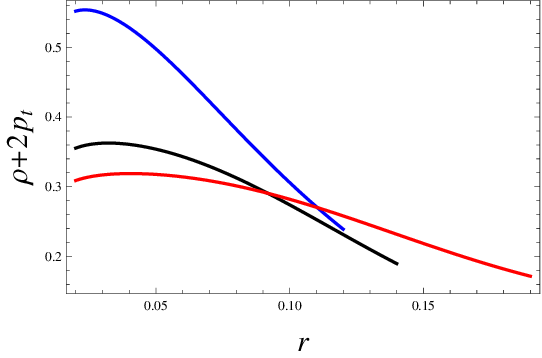,width=0.4\linewidth} \caption{Energy bounds
for $\alpha=0$ (\textcolor{blue}{\textbf{\small $\bigstar$}}), $0.1$
(\textcolor{black}{\textbf{\small $\bigstar$}}) and $0.2$
(\textcolor{red}{\textbf{\small $\bigstar$}}) corresponding to model
I.}
\end{figure}
\begin{figure}\center
\epsfig{file=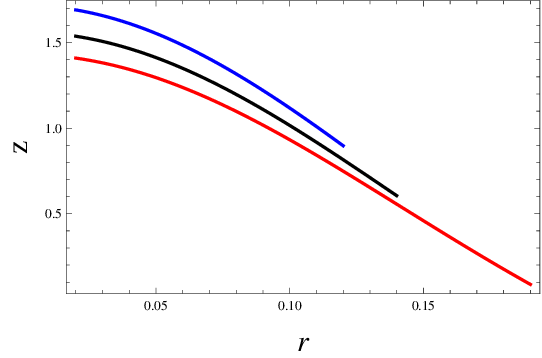,width=0.4\linewidth}\epsfig{file=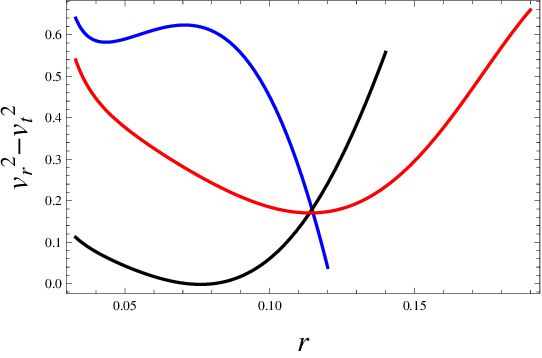,width=0.4\linewidth}
\caption{Redshift and cracking for $\alpha=0$
(\textcolor{blue}{\textbf{\small $\bigstar$}}), $0.1$
(\textcolor{black}{\textbf{\small $\bigstar$}}) and $0.2$
(\textcolor{red}{\textbf{\small $\bigstar$}}) corresponding to model
I.}
\end{figure}

\subsection{Polytropic Model admitting Null Complexity Factor}

In modern research, the polytrope linked with the matter
distribution characterizing anisotropic interior is of significant
importance. The polytropic solutions have been addressed and
explored by several researchers in the context of various
gravitational proposals \cite{12}-\cite{14}. To derive a solution
for the field equations \eqref{g8a}-\eqref{g10a}, we assume an
equation of state representing polytropes and assign a value of zero
to the complexity factor. A concise explanation of the polytropic
model is given in \cite{25}, but we offer a more detailed analysis
of the solution through pictorial representation. The two conditions
discussed above are presented below to continue our analysis as
\begin{align}\label{g55}
p_r=\mathcal{I}\rho^{\eta_3}=\mathcal{I}\rho^{1+\frac{1}{\mathcal{U}}},
\quad \mathcal{Y}_{TF}=0,
\end{align}
where
\begin{itemize}
\item $\mathcal{I}$ refers to the polytropic constant,
\item $\mathcal{U}$ is the symbol for the polytropic index,
\item $\eta_3$ being a polytropic exponent.
\end{itemize}
The limitations defined in Eq.\eqref{g55} yield two non-linear
equations in $\delta_1$ and $\delta_2$, respectively, as
\begin{align}\nonumber
&\frac{e^{-\delta_2}}{2r^2}\big[r \big\{\alpha  \big(r \delta_1
'+4\big) \big(\delta_2 '-\delta_1 '\big)-2 \alpha  r \delta_1 ''+2
\delta_1 '\big\}+2 (2 \alpha -1) \big(e^{\delta_2
}-1\big)\big]-\mathcal{I}\\\nonumber
&\times\big(\frac{e^{-\delta_2}}{2r^2}\big)^{1+\frac{1}{\mathcal{U}}}
\big[r \delta_2 ' \big(2-4 \alpha -\alpha  r \delta_1 '\big)-2 (2
\alpha -1) \big(e^{\delta_2 }-1\big)+\alpha  r \big\{2 r \delta_1
''+\delta_1 '\\\label{g55a} &\times \big(r \delta_1
'+4\big)\big\}\big]^{1+\frac{1}{\mathcal{U}}}=0,\\\nonumber &r
\big[\delta_1 ' \big\{\alpha  r^2 \big(\delta_2 ''-2 \delta_1
''\big)+2 (r-4 \alpha )\big\}-2 r \big\{(1-2 \alpha ) \delta_2
''+\alpha  r \delta_1 ^{(3)}+(5 \alpha +r) \delta_1
''\big\}\\\nonumber &-r \delta_2 '^2 \big(4 \alpha +\alpha  r
\delta_1 '-2\big)-r (3 \alpha +r) \delta_1 '^2+\delta_2 ' \big\{r
\big(3 \alpha r \delta_1 ''+\delta_1 ' \big(7 \alpha +\alpha r
\delta_1 '+r\big)\big)\\\label{g55b} &+2 (6 \alpha
+r-3)\big\}\big]-2 (1-2 \alpha +2 r) \big(e^{\delta_2 }-1\big)=0.
\end{align}
With chosen values of $\mathcal{I} = 0.9$ and $\mathcal{U} = 0.05$,
we solve Eqs.\eqref{g55a} and \eqref{g55b} to calculate $\delta_1$
and $\delta_2$. Through numerical computation of these equations, we
illustrate the graphical trend of the produced metric potentials in
Figure \textbf{5}, presenting a satisfactory profile. Notable, we
are unable to deriving their explicit expressions due to numerically
solving the above two equations. The findings show that
$e^{\delta_1(0)} = c_2$ (a positive constant), and $e^{-\delta_2(0)}
= 1$. For various values of $\alpha$, the surface boundary and
compactness factor are \textcolor{blue} {\begin{itemize}
\item At $\alpha = 0$, we get $\texttt{R}=0.095$, and the compactness is
$$e^{\delta_1(0.095)}=e^{-\delta_2(0.095)}\approx0.21=1-\frac{2\texttt{M}}{\texttt{R}}
\quad \Rightarrow \quad \frac{2\texttt{M}}{\texttt{R}} \approx0.79 <
0.889.$$
\item Setting $\alpha = 0.1$, we have $\texttt{R} = 0.10$, and the related compactness is
$$e^{\delta_1(0.10)}=e^{-\delta_2(0.10)}\approx0.24=1-\frac{2\texttt{M}}{\texttt{R}}
\quad \Rightarrow \quad \frac{2\texttt{M}}{\texttt{R}} \approx0.76 <
0.772.$$
\item At $\alpha = 0.2$, we have $\texttt{R}=0.12$, and the compactness is given by
$$e^{\delta_1(0.12)}=e^{-\delta_2(0.12)}\approx0.29=1-\frac{2\texttt{M}}{\texttt{R}}
\quad \Rightarrow \quad \frac{2\texttt{M}}{\texttt{R}} \approx0.71 <
0.714.$$
\end{itemize}}
\begin{figure}\center
\epsfig{file=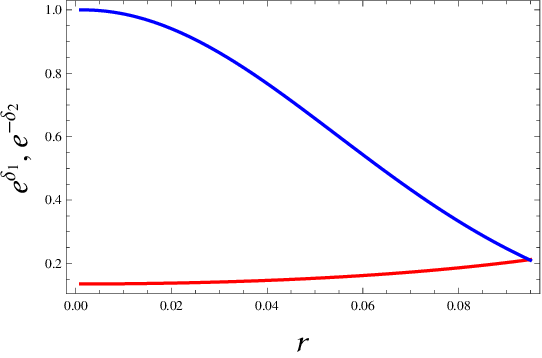,width=0.4\linewidth}\epsfig{file=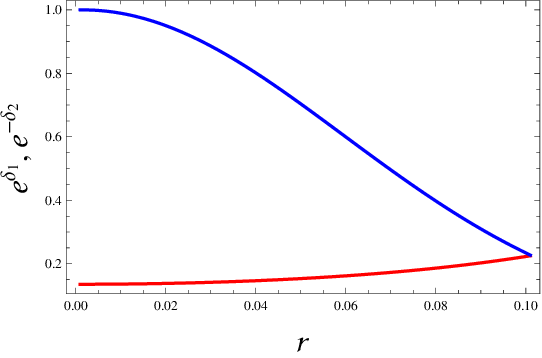,width=0.4\linewidth}
\epsfig{file=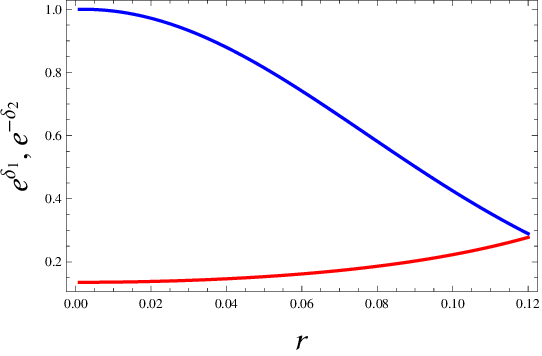,width=0.4\linewidth}
\caption{Metric potentials $e^{\delta_1}$
(\textcolor{red}{\textbf{\small $\bigstar$}}) and $e^{-\delta_2}$
(\textcolor{blue}{\textbf{\small $\bigstar$}}) for $\alpha=0$ (upper
left), $0.1$ (upper right) and $0.2$ (lower) analogous to model II.}
\end{figure}

The profile of the radial/tangential pressures and energy density is
displayed in Figure \textbf{6}, based on the given parametric
values. Their maximum occurs at the core, and the minimum at the
boundary, following a monotonically decreasing trend. The upper left
plot shows that $p_r$ disappears at the interface for every value of
$\alpha$. Figure \textbf{7} displays the plots of all energy
conditions, which are satisfied across all values of $\alpha$. As a
result, they form a solution that is physically admissible and
encompasses ordinary matter.

In the left plot of Figure \textbf{8}, we show the gravitational
redshift for this model, which decreases with increasing $r$. This
factor is given by $z(0.095) = 1.745$, $z(0.10) = 1.432$, and
$z(0.12) = 0.979$ for $\alpha = 0$, $0.1$, and $0.2$, respectively.
Figure \textbf{8} also investigates the stable region, indicating
that cracking is observed only for $\alpha = 0$. This value results
in an unstable system. On the other hand, the remaining two values
result in physically stable interiors. As a result, Rastall theory
offers more favorable outcomes in contrast to GR \cite{40}.
\begin{figure}\center
\epsfig{file=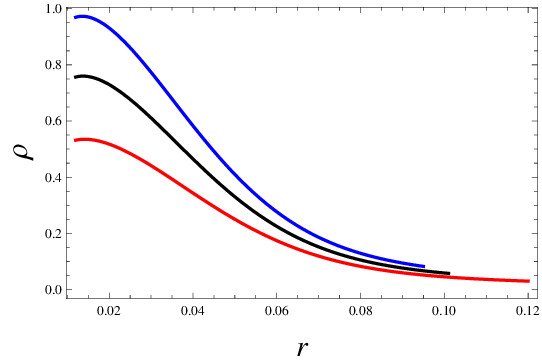,width=0.4\linewidth}\epsfig{file=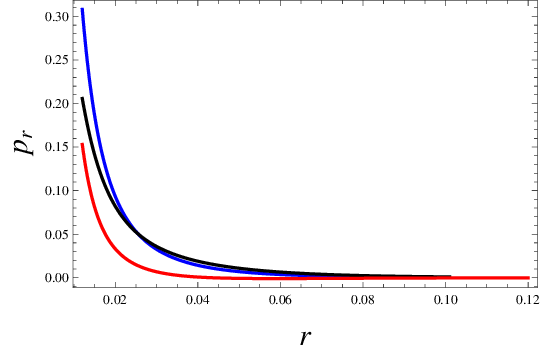,width=0.4\linewidth}
\epsfig{file=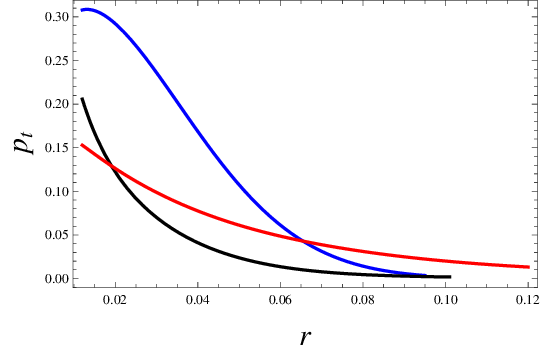,width=0.4\linewidth}\epsfig{file=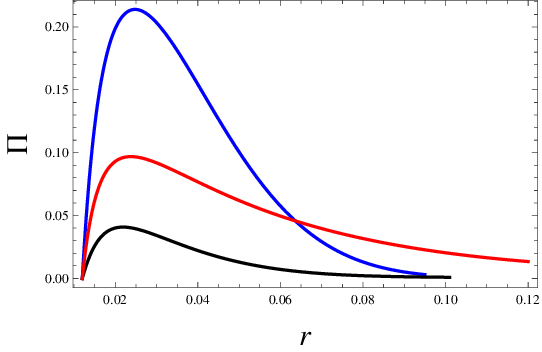,width=0.4\linewidth}
\caption{Governing parameters for $\alpha=0$
(\textcolor{blue}{\textbf{\small $\bigstar$}}), $0.1$
(\textcolor{black}{\textbf{\small $\bigstar$}}) and $0.2$
(\textcolor{red}{\textbf{\small $\bigstar$}}) corresponding to model
II.}
\end{figure}
\begin{figure}\center
\epsfig{file=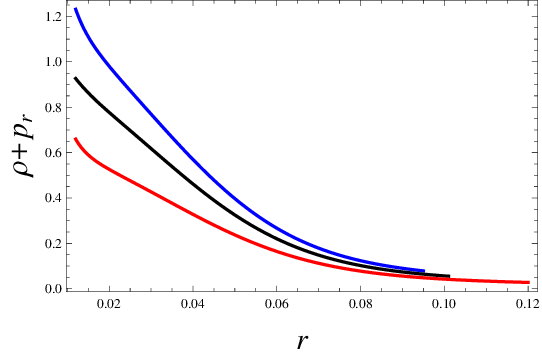,width=0.4\linewidth}\epsfig{file=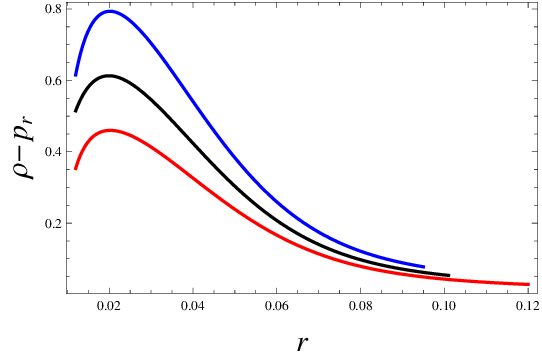,width=0.4\linewidth}
\epsfig{file=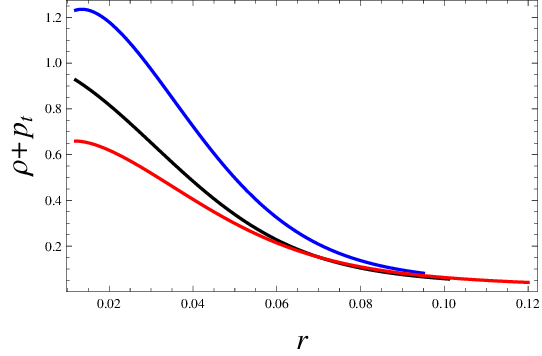,width=0.4\linewidth}\epsfig{file=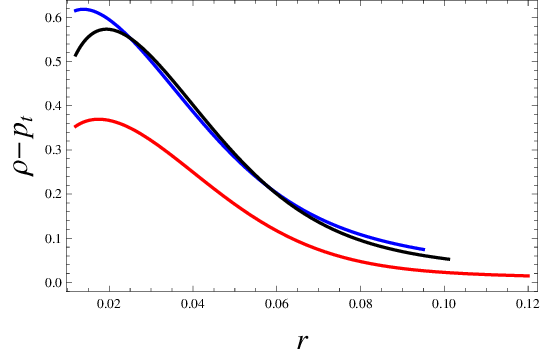,width=0.4\linewidth}
\epsfig{file=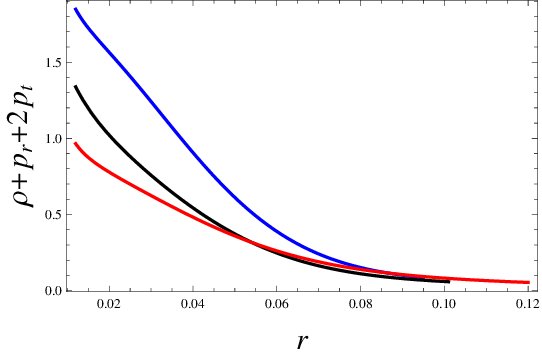,width=0.4\linewidth} \caption{Energy
bounds for $\alpha=0$ (\textcolor{blue}{\textbf{\small
$\bigstar$}}), $0.1$ (\textcolor{black}{\textbf{\small $\bigstar$}})
and $0.2$ (\textcolor{red}{\textbf{\small $\bigstar$}})
corresponding to model II.}
\end{figure}
\begin{figure}\center
\epsfig{file=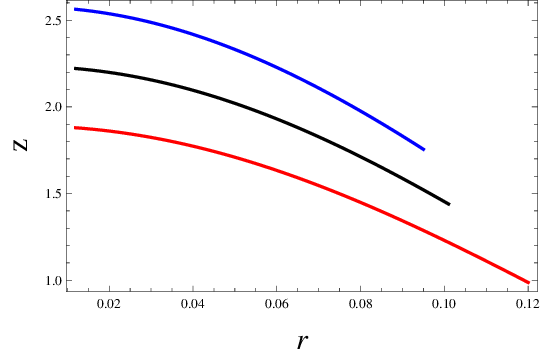,width=0.4\linewidth}\epsfig{file=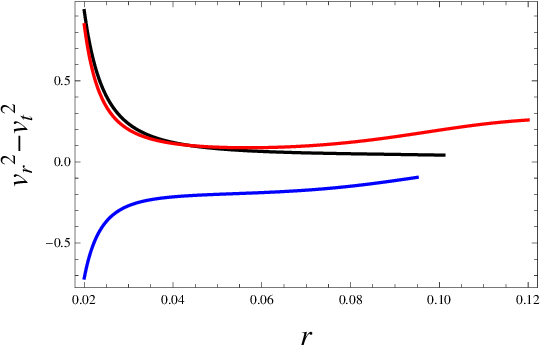,width=0.4\linewidth}
\caption{Redshift and cracking for $\alpha=0$
(\textcolor{blue}{\textbf{\small $\bigstar$}}), $0.1$
(\textcolor{black}{\textbf{\small $\bigstar$}}) and $0.2$
(\textcolor{red}{\textbf{\small $\bigstar$}}) corresponding to model
II.}
\end{figure}

Converting differential equations into the dimensionless form is a
more efficient approach to finding their solution. For this purpose,
we define new variables which are dimensionless as follows
\begin{align}\label{g56}
\rho_c=\frac{p_{rc}}{\tau}, \quad r=\frac{\phi}{\mathcal{B}}, \quad
\mathcal{B}^2=\frac{4\pi\rho_c}{\tau(\mathcal{U}+1)},\\\label{g57}
\Gamma^{\mathcal{X}}=\frac{\rho}{\rho_c}, \quad
\lambda(\phi)=\frac{\mathcal{B}^3m(r)}{4\pi\rho_c},
\end{align}
with $\rho_{c}$ and $p_{rc}$ being the central density and central
radial pressure, respectively. When $r=\texttt{R}$ (or $\phi(r)$
equals $\phi(\texttt{R})$), it holds that
$\Gamma\big(\phi(\texttt{R})\big) = 0$. By combining the above
variables with the mass function \eqref{g12b} and the generalized
evolution equation \eqref{g12d}, the following expression is
obtained
\begin{align}\label{g58}
&\frac{d\lambda}{d\phi}=2\Gamma^\mathcal{U}\phi^2\left(\frac{3\alpha-1}{4\alpha-1}\right)-\frac{2\alpha\phi^2\Pi}{\rho_c(4\alpha-1)}
+\frac{3\alpha\tau\phi^2\Gamma^{1+\mathcal{U}}}{4\alpha-1},\\\nonumber
&\tau(\mathcal{U}+1)\frac{d\Gamma}{d\phi}+\frac{\phi^2\rho_c(1+\tau\Gamma)}{2\mathcal{B}^2}
\bigg[\tau\Gamma^{\mathcal{U}+1}\left(\frac{\alpha-1}{4\alpha-1}\right)+\frac{a\Gamma^\mathcal{U}}{4\alpha-1}
+\frac{2\Pi}{\rho_c(4\alpha-1)}+\frac{\lambda}{\phi^3}\bigg]\\\label{g59}
&+\frac{2\Pi\Gamma^{-\mathcal{U}}}{\phi\rho_c}+\frac{1}{4\alpha-1}\bigg[\alpha
\mathcal{U}\Gamma^{-1}\frac{d\Gamma}{d\phi}-3\alpha\tau(\mathcal{U}+1)\frac{d\Gamma}{d\phi}
+\frac{2\alpha\Gamma^{-\mathcal{U}}}{\rho_c}\frac{d\Pi}{d\phi}\bigg]=0.
\end{align}

As the two differential equations \eqref{g58} and \eqref{g59}
involve the variables $\Gamma,~\lambda$, and $\Pi$, to uniquely
identify them, an additional condition is needed. In order to
satisfy this condition, we select $\mathcal{Y}_{TF} = 0$, which,
when rewritten using the variables \eqref{g56} and \eqref{g57}, is
given as follows
\begin{align}\nonumber
&6\Pi+2\phi\frac{d\Pi}{d\phi}+2\phi\Pi\frac{d\delta_2}{d\phi}=\frac{\alpha
\mathcal{B}^2e^{-\delta_2}}{\phi}\left(\frac{\varpi}{\phi}+\frac{d\varpi}{d\phi}\right)
+2\rho_c\Gamma^{\mathcal{U}-1}\left(\mathcal{U}\phi\frac{d\Gamma}{d\phi}+2\right)+2\\\label{g60}
&\times\left(1+\phi\frac{d\delta_2}{d\phi}\right) \left(\rho_c
\Gamma^{\mathcal{U}}+\frac{2\alpha \mathcal{B}^2}{\phi^2}\right)
-\frac{4\rho_c\Gamma^\mathcal{U}(3\alpha-1)}{4\alpha-1}-\frac{4\alpha}{4\alpha-1}(3\tau
\Gamma^{1+\mathcal{U}}-2\Pi),
\end{align}
where
$$\varpi=\phi\left(\phi\frac{d\delta_1}{d\phi}+4\right)\left(\frac{d\delta_2}{d\phi}-\frac{d\delta_1}{d\phi}\right)
-2\phi^2\frac{d^2\delta_1}{d\phi^2}-4.$$

As $\delta_1$ and $\delta_2$ are already known, the number of
equations is equal to those of unknowns. As a result, a numerical
solution is found that is based on the initial conditions set by
\begin{align}\nonumber
\Pi(\phi)|_{\phi=0}=0, \quad \Gamma(\phi)|_{\phi=0}=1, \quad
\lambda(\phi)|_{\phi=0}=0.
\end{align}
For $\tau = 0.1$, we investigate the graphical behavior of the above
mentioned unknowns against $\phi$ by altering the values of
$\mathcal{U}$. The energy density, shown in Figure \textbf{9} (upper
left), peaks at $\phi = 0$ and diminishes outward, indicative of the
well-behaved nature of polytropes. As illustrated in the upper right
plot, the corresponding mass increases steadily, showing an inverse
relationship with $\mathcal{U}$.
\begin{figure}\center
\epsfig{file=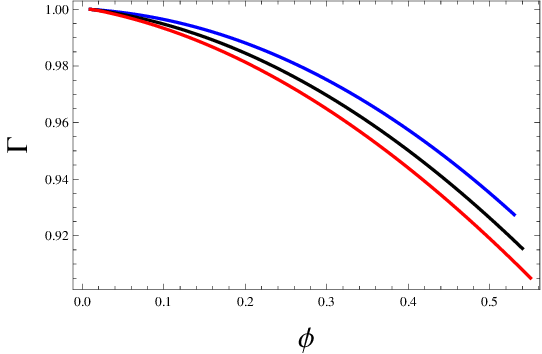,width=0.4\linewidth}\epsfig{file=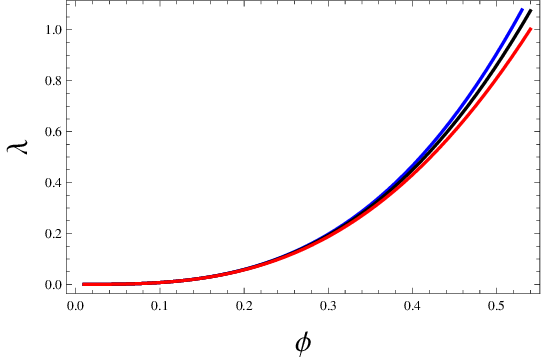,width=0.4\linewidth}
\epsfig{file=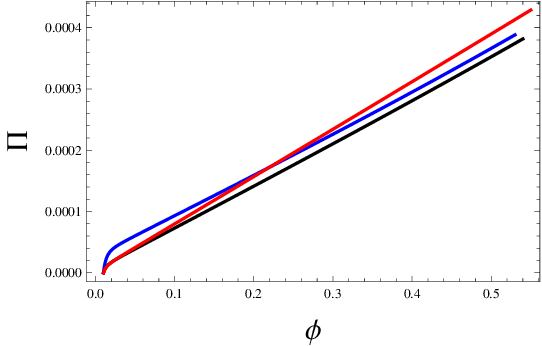,width=0.4\linewidth}
\caption{Profile of $\Gamma,~\lambda$ and $\Pi$ for
$\mathcal{U}=0.06$ (\textcolor{blue}{\textbf{\small $\bigstar$}}),
$0.07$ (\textcolor{black}{\textbf{\small $\bigstar$}}) and $0.08$
(\textcolor{red}{\textbf{\small $\bigstar$}}) corresponding to model
II.}
\end{figure}

\subsection{Model admitting Non-local Equation of State and $\mathcal{Y}_{TF}=0$}

Hernandez and Nunez \cite{ai} proposed a constraint that relates the
radial pressure to both the energy density and an integral term, the
latter of which is included in the complexity factor \eqref{g30} and
is known as the non-local equation of state. This equation together
with the complexity-free condition are given by
\begin{align}\label{g62}
p_r=\rho-\frac{2}{r^3}\int_0^r\rho\bar{r}^2d\bar{r}+\frac{c_3}{2\pi
r^3}, \quad \mathcal{Y}_{TF}=0,
\end{align}
where $c_3$ stands for a real-valued constant. To eliminate the
possibility of the singularity at the star's core, we define this
constant as zero. The equation shown above (left) can be restated by
connecting it with the interior mass \eqref{g12b} as
\begin{align}\label{g63}
&4 \pi  r \big[\big(4 \alpha +\alpha  r \delta_1 '-1\big)
\big(\delta_2 '-\delta_1 '\big)-2 \alpha  r \delta_1 ''\big]+\big\{8
\pi \big(2 \alpha -1\big)+1\big\} \big(e^{\delta_2 }-1\big)=0.
\end{align}
Nevertheless, the condition $ \mathcal{Y}_{TF} = 0 $ continues to
apply as specified in Eq.\eqref{g55b}. As a result, both of these
equations are adequate to find the solution to
Eqs.\eqref{g8a}-\eqref{g10a}. It is important to point out that
Eqs.\eqref{g55b} and \eqref{g63} include higher-order expressions
involving the metric components, necessitating a numerical solution,
as previously carried out in the earlier models.

As seen in Figure \textbf{10}, the profiles of $e^{\delta_1}$ and
$e^{-\delta_2}$ are non-singular, positive, and finite, signifying
that these potentials are physically acceptable. We also establish
that $e^{\delta_1(0)} = c_4~(>0)$ and $e^{-\delta_2(0)} = 1$ at the
center, with both functions converging to a single point known as
radius. The summary is shown below \textcolor{blue} {\begin{itemize}
\item When $\alpha = 0$, we obtain $\texttt{R} = 0.11$, and the compactness takes the following value
$$e^{\delta_1(0.11)}=e^{-\delta_2(0.11)}\approx0.25=1-\frac{2\texttt{M}}{\texttt{R}}
\quad \Rightarrow \quad \frac{2\texttt{M}}{\texttt{R}} \approx0.75 <
0.889.$$
\item When $\alpha = 0.1$, we obtain $\texttt{R} = 0.12$, and the compactness takes the following
value
$$e^{\delta_1(0.12)}=e^{-\delta_2(0.12)}\approx0.28=1-\frac{2\texttt{M}}{\texttt{R}}
\quad \Rightarrow \quad \frac{2\texttt{M}}{\texttt{R}} \approx0.72 <
0.772.$$
\item When $\alpha = 0.2$, we obtain $\texttt{R} = 0.16$, and the compactness is expressed as
$$e^{\delta_1(0.16)}=e^{-\delta_2(0.16)}\approx0.48=1-\frac{2\texttt{M}}{\texttt{R}}
\quad \Rightarrow \quad \frac{2\texttt{M}}{\texttt{R}} \approx0.52 <
0.714.$$
\end{itemize}}
\begin{figure}\center
\epsfig{file=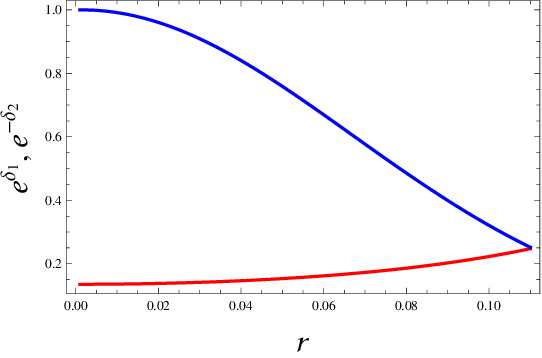,width=0.4\linewidth}\epsfig{file=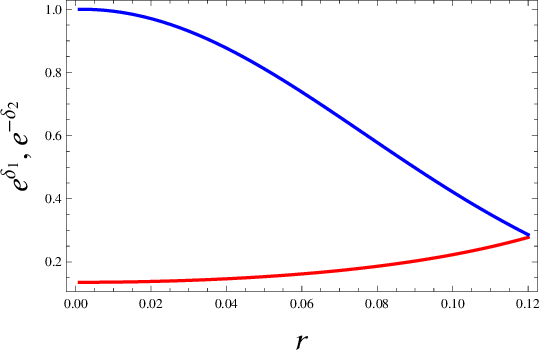,width=0.4\linewidth}
\epsfig{file=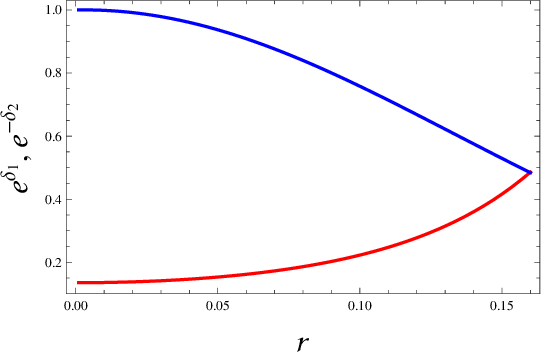,width=0.4\linewidth}
\caption{Metric potentials $e^{\delta_1}$
(\textcolor{red}{\textbf{\small $\bigstar$}}) and $e^{-\delta_2}$
(\textcolor{blue}{\textbf{\small $\bigstar$}}) for $\alpha=0$ (upper
left), $0.1$ (upper right) and $0.2$ (lower) analogous to model
III.}
\end{figure}

Figure \textbf{11} depicts the fluid triplet's acceptable nature,
with a positive maximum at the core that decreases as it extends to
the boundary. Significantly, the difference between the two
pressures gives rise to positive anisotropy, differing from the
outcome in $f(R)$ theory \cite{20}, as shown in the lower right
plot. As can be seen in Figure \textbf{12}, the energy bounds are
adhered to, validating the applicability of the model with usual
matter. As shown in Figure \textbf{13} (left), the redshift
decreases with increasing $r$, achieving values of $z(0.11) =
1.197$, $z(0.12) = 1.008$, and $z(0.16) = 0.613$ for $\alpha = 0$,
$0.1$, and $0.2$ at the boundary. In the right plot, the cracking
condition is positive within the $[0,1]$ range throughout the
interior, except for the case of $\alpha = 0.2$. Thus, this solution
analogous to the constraints \eqref{g34} and \eqref{g62} is stable
for $\alpha=0$ and $\alpha=0.1$, which is consistent with \cite{40}.
\begin{figure}\center
\epsfig{file=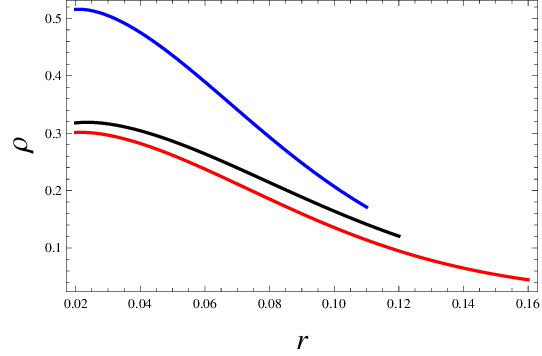,width=0.4\linewidth}\epsfig{file=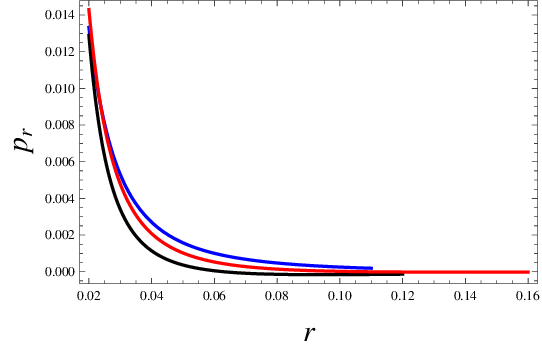,width=0.4\linewidth}
\epsfig{file=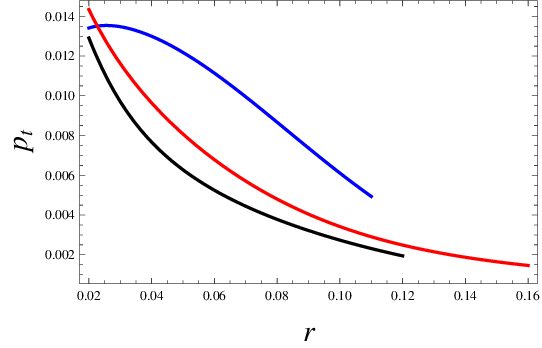,width=0.4\linewidth}\epsfig{file=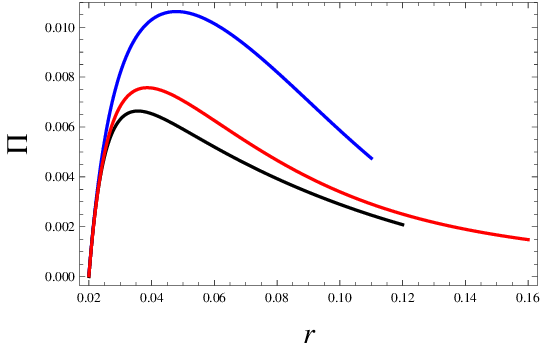,width=0.4\linewidth}
\caption{Governing parameters for $\alpha=0$
(\textcolor{blue}{\textbf{\small $\bigstar$}}), $0.1$
(\textcolor{black}{\textbf{\small $\bigstar$}}) and $0.2$
(\textcolor{red}{\textbf{\small $\bigstar$}}) corresponding to model
III.}
\end{figure}
\begin{figure}\center
\epsfig{file=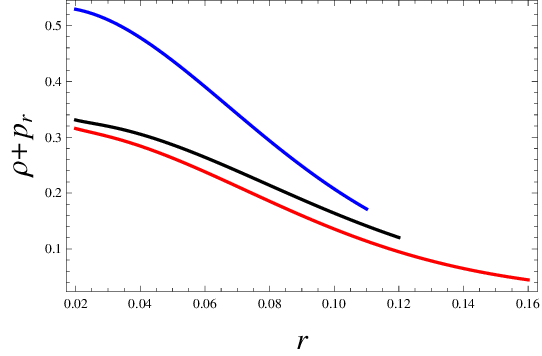,width=0.4\linewidth}\epsfig{file=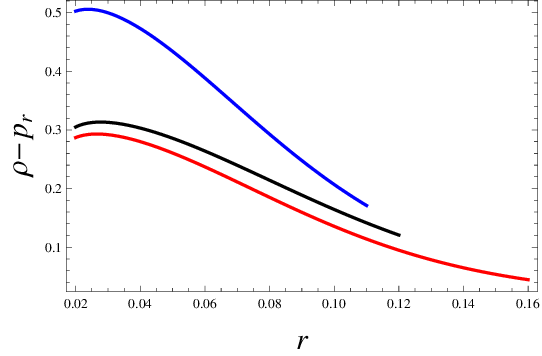,width=0.4\linewidth}
\epsfig{file=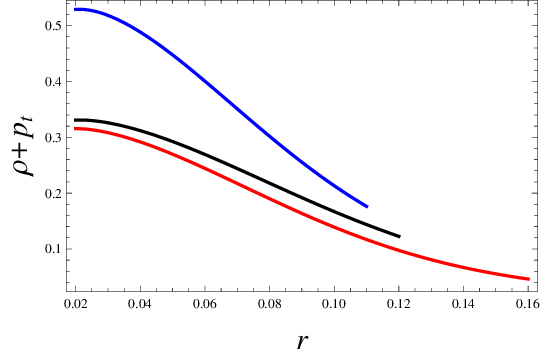,width=0.4\linewidth}\epsfig{file=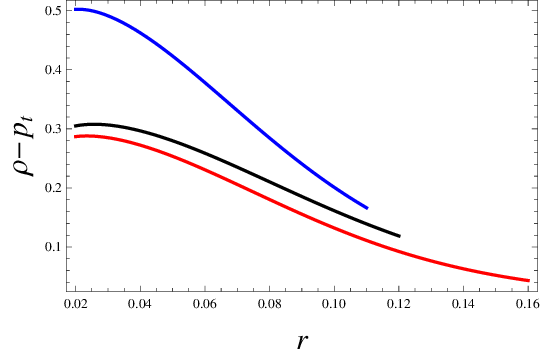,width=0.4\linewidth}
\epsfig{file=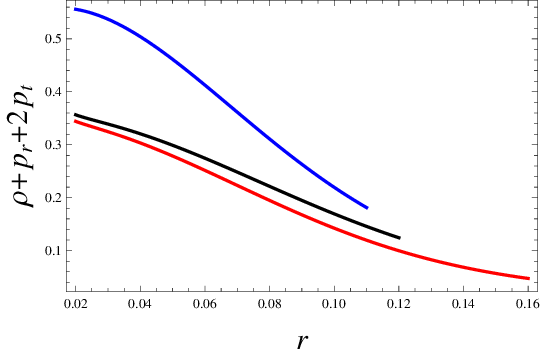,width=0.4\linewidth} \caption{Energy
bounds for $\alpha=0$ (\textcolor{blue}{\textbf{\small
$\bigstar$}}), $0.1$ (\textcolor{black}{\textbf{\small $\bigstar$}})
and $0.2$ (\textcolor{red}{\textbf{\small $\bigstar$}})
corresponding to model III.}
\end{figure}
\begin{figure}\center
\epsfig{file=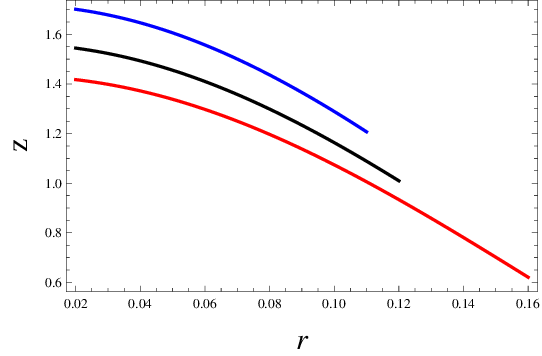,width=0.4\linewidth}\epsfig{file=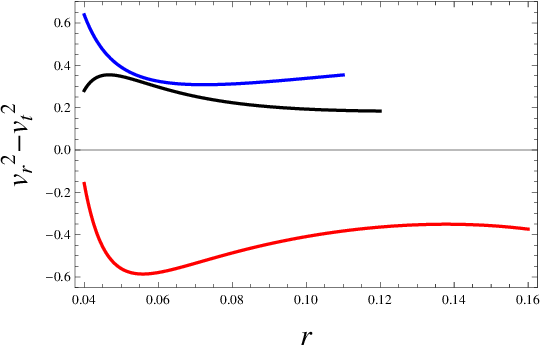,width=0.4\linewidth}
\caption{Redshift and cracking for $\alpha=0$
(\textcolor{blue}{\textbf{\small $\bigstar$}}), $0.1$
(\textcolor{black}{\textbf{\small $\bigstar$}}) and $0.2$
(\textcolor{red}{\textbf{\small $\bigstar$}}) corresponding to model
III.}
\end{figure}

\section{Conclusions}

This paper is aimed at investigating different solutions to the
modified field equations in the non-conserved Rastall theory. In
order to accomplish this, a static sphere has been considered, and
the governing equations and the generalized evolution equation were
calculated. An expression for the spherical mass function has also
been derived in terms of the geometric and matter quantities. The
Schwarzschild metric has then been assumed as the exterior geometry
that facilitated the determination of the compactness factor. As
part of the investigation, the curvature tensor has been
orthogonally to obtain four distinct scalars which are connected to
various physical factors governing the interior of a compact object.
One of these factors, indicated by $ \mathcal{Y}_{TF} $, was found
to possess density inhomogeneity, anisotropy, and modified
corrections, making it the most suitable candidate of the complexity
factor for the fluid distribution under consideration, as indicated
by Herrera \cite{25}.

The system of equations \eqref{g8a}-\eqref{g10a} involves five
independent variables: the fluid's triplet and the metric
potentials. To aid in solving this, appropriate constraints have
been introduced. The complexity-free condition, detailed in
Eq.\eqref{g34}, serves as the first constraint. To further constrain
the system, three conditions have been imposed: $p_r=0$, a
polytropic as well as a non-local equation of state, forming the
second condition along with $\mathcal{Y}_{TF}=0$ and yielding a
triplet of different models. The problem of the involvement of
higher-order terms related to the geometric sector $\delta_1,
\delta_2$ was tackled through numerical integration of the
differential equations, taking into account feasible initial
conditions for each scenario. The next step involves detailing
several physical requirements, including gravitational redshift,
compactness, and stability conditions, that are essential for
constructing realistic compact models.

The fluid sector including the pressure components and energy
density associated with all three solutions exhibit suitable
behavior, peaking at $r=0$ and decreasing as the radius grows. It is
found that both the compactness and the gravitational redshift are
within the acceptable range. The viability criterion has been met by
all solutions for all choices of $\alpha$, as confirmed by the
fulfillment of energy conditions (Figures \textbf{3},~\textbf{7},
and \textbf{12}). It is worth exploring the occurrence of cracking
in the interior of our models, and we highlight the summary as
follows.
\begin{itemize}
\item The cracking condition is fulfilled only by the solution
analogous to $p_r=0$, regardless of the values selected for the
Rastall parameter (Figure \textbf{4}).
\item The stability for the second solution is restricted to
$\alpha=0.1,~0.2$. However, for $\alpha=0$, the model becomes
unstable, which proves the superiority of the Rastall theory over GR
(Figure \textbf{8}).
\item For the third solution, stability is seen only for
$\alpha=0,~0.1$, as cracking occurs for the remaining value (Figure
\textbf{13}).
\end{itemize}

It is crucial to mention that all the developed solutions are
consistent with the results found in GR \cite{40}. It is noted that
the second and third models diverge from the charged scenario
\cite{30}.
\\\\
\textbf{Data Availability:} No additional data were analyzed or
created as part of this study.

\end{document}